\documentclass{iaus}
\usepackage{graphicx}

\title[Generation of Potential Vorticity in Planet-Disk Systems] 
{Baroclinic Generation of Potential Vorticity in an Embedded Planet-Disk System}

\author[Ji et al.]   
{Jianghui JI$^{1}$, Shangli OU$^2$  \and   Lin LIU$^3$}

\affiliation{$^1$Purple  Mountain  Observatory, Chinese  Academy
of  Sciences,  Nanjing  210008, China \break email: jijh@pmo.ac.cn\\[\affilskip]
$^2$High Performance Computing, Center for Computation and
Technology / Information Technology Services, Louisiana State
University, Baton Rouge, LA 70803 \break email: ou@cct.lsu.edu\\[\affilskip]
$^3$Department of Astronomy,  Nanjing University, Nanjing  210093,
China }

\pubyear{2007}
\volume{249}  
\pagerange{1--5}

 \jname{Exoplanets: Detection, Formation and
Dynamics} \editors{Sylvio Ferraz-Mello, Yi-Sui Sun \& Ji-lin Zhou,
eds.}
\begin{document}

\maketitle

\begin{abstract}
We use a multi-dimensional hydrodynamics code to study the
gravitational interaction between an embedded planet and a
protoplanetary disk with emphasis on the generation of vortensity
(Potential Vorticity or PV) through a Baroclinic Instability. We
show that the generation of PV is very common and effective in
non-barotropic disks through the Baroclinic Instability, especially
within the coorbital region. Our results also complement previous
work that non-axisymmetric Rossby-Wave Instabilities (RWIs, Lovelace
et al. 1999) are likely to develop at local minima of PV
distribution that are generated by the interaction between a planet
and an inviscid barotropic disk. The development of RWIs results in
non-axisymmetric density blobs, which exert stronger torques onto
the planet when they move to the vicinity of the planet. Hence,
large amplitude oscillations are introduced to the time behavior of
the total torque acted on the planet by the disk. In current
simulations, RWIs do not change the overall picture of inward
orbital migration but cause a non-monotonic behavior to the
migration speed. As a side effect, RWIs also introduce interesting
structures into the disk. These structures may help the formation of
Earth-like planets in the Habitable Zone or Hot Earths interior to a
close-in giant planet.

\end{abstract}

\firstsection 

\section{Introduction}
The standard core-accretion theory (Safronov 1969; Lissauer 1993)
suggests that the formation of planets in circumstellar disks around
T Tauri stars consists of the formation of planetesimals via
collisions/coalitions of dust grains in the early stage and then
gravitationally accretion after they accumulate enough mass. The
lifetime of this T Tauri star phase is estimated to be short
($\lesssim 10^7$ years). In order for cores of protoplanets with
Jupiter mass to accumulate enough material in the T Tauri stage, it
is thought that their cores have to form outside the so-called ``ice
line" located far away from the central star (typically beyond
$\sim$ 4 AU, see Ida \& Lin 2004) so that the temperature is low
enough to allow the condensation of gas materials to solid ice.
These additional solid grains helps to increase the dust coagulation
speed and shorten the time needed to form the embryos of
protoplanets.  From the observational side, recent discoveries of
extrasolar planets show that a large number of the host stars are
surrounded by "hot Jupiters" and close-in super Earths. Around
$80\%$ of the extrasolar planets are in orbits with semi-major axes
in the range $0.01 \lesssim a \lesssim 2.5$ AU, and $\sim$ 25\% of
the total population are short-period planets with $a \lesssim 0.1$
AU. This has brought one of the most interesting puzzles to
theorists: if protoplanets had formed in a disk region beyond $\sim$
4 AU from the central star, how did the observed extrasolar planets
end up with orbits that are so close to their host stars? If the
standard theory for the formation of protoplanetary cores holds
(Pollack et al. 1996; Ida  \& Lin 2004), then the giant planets
(like hot-Jupiters) must have undergone  inward orbital migration to
their current locations.

Goldreich \& Tremaine (1978) and many other authors studied the
gravitational interaction between a gaseous protoplanetary disk and
an embedded planet (or satellite). The planet causes the formation
of spiral waves inside the disk at the Lindblad resonances; as a
result of the density asymmetry induced by spiral waves, the inner
disk exerts a positive gravitational torque onto the planet and the
outer disk exerts a negative gravitational torque onto the planet.
According to analytical analysis, the overall torque is generally
negative and, hence, forces the planet to migrate inward.  Ward
(1997) pointed out that depending on the mass of the protoplanet two
major kinds of migration exist. In the so called Type I migration,
the planet's mass is small and  the response of the disk is linear;
the migration speed is very fast so that the migration timescale is
as short as $\sim 10^4$ yr for a planetary core of 10 $M_{\oplus}$
in a minimum mass solar nebula at $\sim 5.2$ AU with sufficient
viscosity (see Figure 14 of Ward 1997). In Type II migration, the
protoplanet is massive enough to open a gap inside the disk, and
migrates on a much longer viscous timescale. Besides classical
analytical analysis, many groups have studied the nonlinear
evolution of a disk-planet system using numerical multi-dimensional
hydrodynamics (Kley 1999; Nelson et al. 2000). These numerical
simulations showed that the nonlinear evolution of the orbital
migration of a planet inside a disk agrees with linear analysis in a
qualitative manner. Masset (2001) and  Masset \& Papaloizou (2003)
suggested that the corotation torque originated from the coorbital
region may play a very different role from that of Lindblad torques;
this leads to a third kind of migration referred to as Type III
migration, in which the migration happens on a timescale as short as
a few tens of orbits and can be directed outward in some cases.

Klahr \& Bodenheimer (2003) described a Baroclinic Instability in
non-barotropic disks that may contribute to vorticity and global
turbulence, then argued that strong vorticities may contribute to
rapid formation of Jupiter-size gas planet (Klahr \& Bodenheimer
2006). Koller et al. (2003) and Li et al. (2005) showed that the
so-called RWIs may develop at the local minima of PV, or vortensity
(defined as the ratio between local vorticity and surface density),
in an inviscid disk with initially uniform PV distribution. Li et
al. (2005) showed that non-axisymmetric RWIs lead to the formation
of vorticies and density blobs, which exert stronger torque onto the
planet when they travel around its vicinity and bring large
oscillations to the total torque acted on the planet. They further
argued that this mechanism may be possible to change the direction
of the migration. Non-axisymmetric RWIs are also relevant to
evolution of a single disk (Li et al. 2000) and stellar models with
strong differential rotation (Ou \& Tohline 2006). In this work, we
study the generation of PV through baroclinic effect and subsequent
development of RWIs in systems consisting of a planet and a
non-barotropic disk.

\section{Basic Equations, Methods and  Initial  Setup}
To investigate the interaction between a disk and an embedded planet
requires coupling hydrodynamics and orbital dynamics together. Here,
we follow Nelson et al. (2000) and many previous investigations to
reduce the problem to a two-dimensional (2D) one since the disk
thickness is of the order of or smaller than the planetary Hill
radius. Three dimensional (3D) investigations will be postponed to
future. The fluid motion inside the disk is described by the
vertically integrated continuity equation (2.1), radial and
azimuthal components of the Navier-Stokes equation (2.2) and (2.3),
\begin{eqnarray}
   \frac{\partial \Sigma} {\partial t} + \nabla \cdot (\Sigma \vec{v}) &=& 0 \label{conteq}\\
   \frac{\partial(\Sigma v_r)} {\partial t} + \nabla \cdot (\Sigma v_r \vec{v})
      &=& \frac{\Sigma v^2_{\phi}} {r} - \frac{\partial P}{\partial r}
         - \Sigma \frac{\partial \Phi}{\partial r} + f_r  \label{NS_eq_r}\\
   \frac{\partial(\Sigma v_{\phi})} {\partial t} + \nabla \cdot (\Sigma v_{\phi} \vec{v})
      &=& -\frac{\Sigma v_r v_{\phi}} {r} - \frac{1}{r} \frac{\partial P}{\partial \phi}
         - \frac{\Sigma}{r} \frac{\partial \Phi}{\partial \phi} + f_{\phi} \label{NS_eq_p} \,,
\end{eqnarray}
where $\Sigma$ is disk surface density, $\vec{v}$ is two fluid
velocities, $P$ is vertically integrated pressure, $f_r$ and
$f_{\phi}$ are two components of viscous forces, and $\Phi$ is the
gravitational potential felt by fluid elements. Details regarding
viscous terms can be found in Nelson et al. (2000). The EOS of the
disk fluid is considered as locally isothermal (Nelson et al. 2000)
as given by $P = c_s^2 \Sigma$, where the local isothermal sound speed
is $c_s = \frac{H}{r} \sqrt{GM_*/r}$ with disk aspect ratio
$H/r=0.05$. In order to compare with globally isothermal situation,
we also carried out one run with a uniform value of $c_s=\frac{H}{r} \sqrt{GM_*/r_p}$,
where $r_p=1$.

We further simplify our study to non-self-gravitational systems, in
which the self-gravity of the fluid is not taken into account for
the fluid motion; hence, $\Phi = \Phi_* + \Phi_p$, where $\Phi_*$ is
the potential field of the central star and $\Phi_p$ is the
potential field of the planet, which is given by $\Phi_p=- M_p
/\sqrt{r^2+\epsilon^2}$, where $M_p$ is the planet mass and
$\epsilon$ is taken to be 0.2 times the Roche Lobe of the planet.
The initial disk model has Keplerian rotational profile and uniform
density, which results in an initial radial PV profile $\xi(r)$ that
is proportional to $r^{-\frac{3}{2}}$. The value of density and
viscosity are chosen to follow those specified in de Val-Borro et
al. (2006). To handle the hydrodynamics part, we adopted a legacy
code developed by the astrophysical group at Louisiana State
University to study star formation (Tohline 1980). The code is
explicit and 2nd order in both space and time. It splits the source
term and advection term in a manner similar to Zeus (Stone \& Norman
1992). Other features implemented include Van Leer upwind  scheme,
artificial viscosity to handle shock, and, staggered cylindrical
grids. The code is originally three-dimensional, but adapted to 2D
in this work.  At the boundary of our computational grids, mass is
allowed to flow off the grids but no inflow is allowed. We also
implemented the wave-killing boundary condition specified in de
Val-Borro et al. (2006). The planet is put on a fixed orbit at $r=1$
AU in most of our simulations. We also allowed the planet to move in
some runs to study the effect of RWIs on its migration. The units
adopted are the following: the gravitational constant $G=1$, length
unit is 1 AU, and $M_*+M_p=1$. Some brief results follow in the next
section.

\begin{figure}


\includegraphics[height=1.7in, width=2.2in]{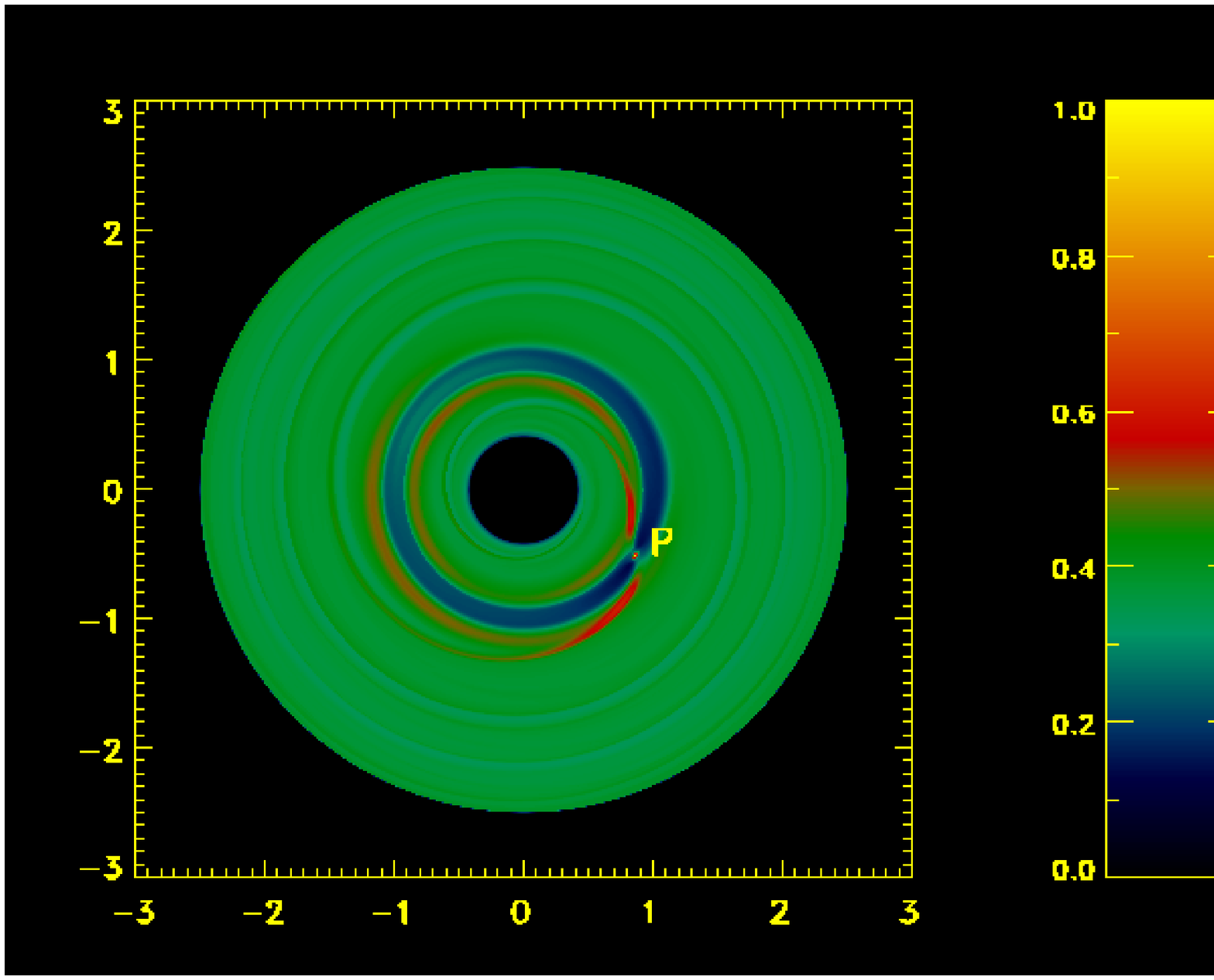}
\centering \hspace{0.5in}
\includegraphics[height=1.7in, width=2.2in]{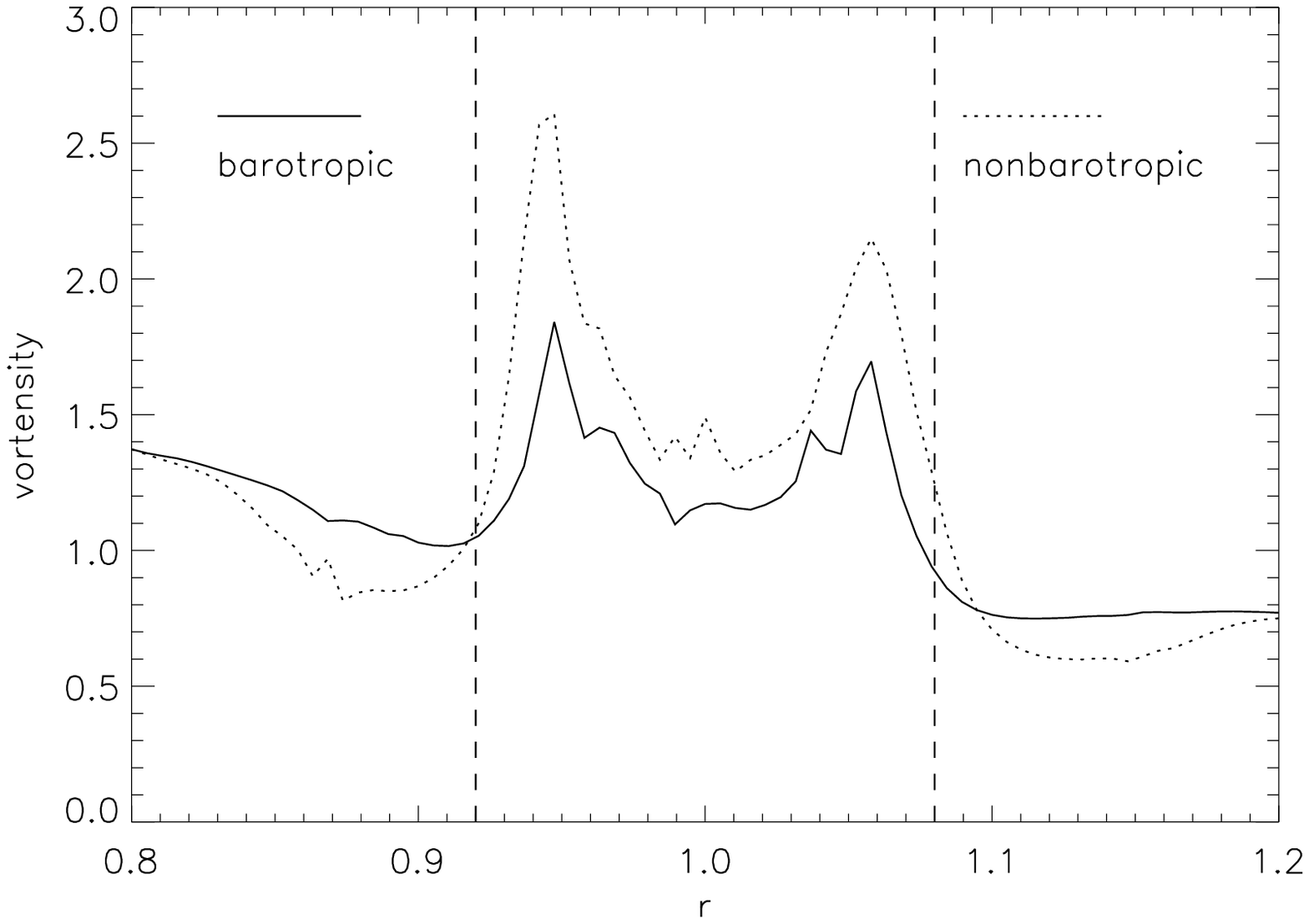}
  \caption{Simulations for an embedded Neptune-mass planet in an inviscid disk.
  \textit{Left panel}: (1a) Color map of the disk surface density
   distribution in linear scale $t \approx 150$ orbits for a resolution run (400 x 1600).
   A letter "P" is labeled next to the location of the planet.
   The color bar represents relative rather than absolute values.
  \textit{Right panel}: (1b) Comparison of azimuthally-averaged radial PV
   profile between two disk-planet systems with a planet embedded in
   a locally isothermal disk (non-barotropic) and an isothermal disk (barotropic).
   Vertical dashed lines illustrate edges of the horseshoe region.
   The planet is located at $r=1$.
} \label{fig1}
\end{figure}

\section{Simulation results}
We present simulation results for systems with a Neptune-mass planet
embedded in either a locally isothermal disk (non-barotropic) or an
isothermal disk (barotropic). Fig.1a illustrates a linear color map
of surface density distribution on a polar plot for a locally
isothermal disk at $t \approx 150$ orbits. Focusing on the formation
of high density areas (red/brown regions), we observe not only red
Lindblad spiral arms, but also, other non-axisymmetric high density
structures at different locations: the inner edge of the outer disk
(brown arc-shaped region centered around 7 o'clock) , the outer edge
of the inner disk (red/brown arc-shaped region around its edge);
interestingly, the density inside the gap is no longer axisymmetric
any more, as suggested by the light blue region around 10 oclock.
The radial locations of these non-axisymmetric structures match
exactly with the local minima of the PV distribution (see Fig.4 of
Ou et al. 2007), which is consistent with previous studies on RWIs
(Li et al. 2000) that RWIs are capable of introducing
non-axisymmetric density distribution in a disk. The total torque
acted on the planet is negative in average; however, as time goes
on, it varies greatly in amplitude (see Ou et al. 2007 for more
details). This is because those high density blobs formed through
RWIs exert a stronger torque on the planet when they are traveling
in the vicinity of the planet, hence, large amplitude of
oscillations show up in the time evolution of the total torque. Our
results on the temporal oscillating behavior of the total torque is
consistent with those in Li et al. (2005).

In general, we expect vorticity (hence, PV) to be generated around
the shock outside coorbital region, where density gradient and
pressure gradient are not aligned to each other. However, detailed
analysis shows that PV is generated both near shocks and around the
planet in the coorbital region. To understand the mechanism that
generates PV within the coorbital region, we note that our disk
models have a non-barotropic EOS with a radial variation of sound
speed $c_s$, which brings in misalignment of pressure gradient and
density gradient wherever azimuthal density gradient appears
($\nabla c_s \times \nabla \Sigma \ne 0$). This misalignment acts as
a source term for the generation of vorticity within the coorbital
region, where no shock but strong azimuthal density gradient
presents. Therefore, the large density depression in the same region
naturally gives birth to PV increment. Fig.1b shows a comparison of
azimuthally averaged radial PV profiles for two runs at $t \approx
60$ orbits: the disk in one run has an isothermal EOS (barotropic)
and the other disk has a locally isothermal EOS (non-barotropic).
The PV peaks in the non-barotropic disk is significantly higher than
those of barotropic disk. Furthermore, a peak located around the
planet ($r=1$) shows up in the non-barotropic disk. To further
examine if our results are resolution-dependent, we carried out a
run with resolution $800\times3200$ for $\sim 70$ orbits. The
simulation exhibits PV maxima within the coorbital region as well
(see Ou et al. 2007). As a further evidence that PV is generated
around where azimuthal density gradient exists, Figure 2a and 2b
illustrate the distribution of PV increment and azimuthal density
gradient, respectively, for the run with resolution $800 \times
3200$ at $ t \sim 70$ orbits. For a better view, they are zoomed in
the neighborhood of the planet. It is observed that PV is generated
within the Roche lobe, where strong azimuthal density gradient,
instead of shocks, exists. Such a generating mechanism of PV inside
a planet's Roche lobe has the same origin as baroclinic instability
discussed in Klahr \& Bodenheimer (2003), which contributes to
global turbulence within the disk. On the other hand, we also
observe that the oscillations of the torque acted on a Neptune-mass
planet appear much earlier in our simulations compared to previous
studies, which suggests that PV generation is even more effective
and common in non-barotropic disk models (see also Ou et al. 2007).

As shown in Figure 2b, the azimuthal density gradient flips the sign
across the planet; thus, we expect the PV increment also flip the
sign there. However, the PV change within the Roche-lobe shown in
Figure 2a is always positive, this is possibly due to PV mixture
along librating stream lines in the horse region where fluid
elements make the U-turn. A detailed and rigorous analysis of fluid
dynamics and PV mixture around a Neptune-mass planet embedded in a
non-barotropic disk will be carried out in a separate investigation.

In our simulations of a freely moving Neptune-mass planet (Ou et al.
2007), RWIs do not change the overall picture of inward migration of
a Neptunian planet; but they have significant influence on the
torque exerted on the planet and make the migration speed of the
planet non-monotonic.

\begin{figure}
\includegraphics[height=1.7in, width=2.2in]{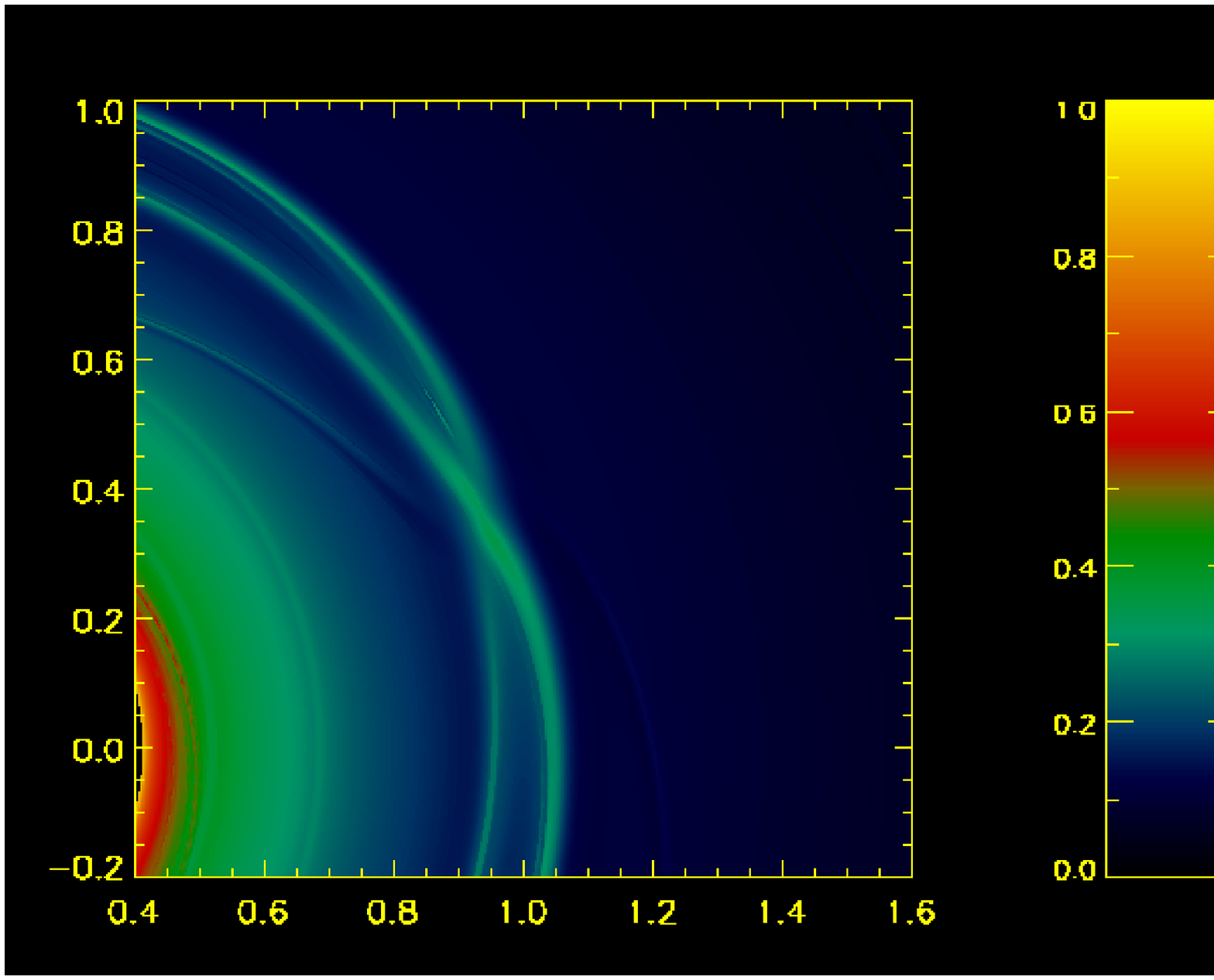}
\hspace{0.4in}
\includegraphics[height=1.7in, width=2.2in]{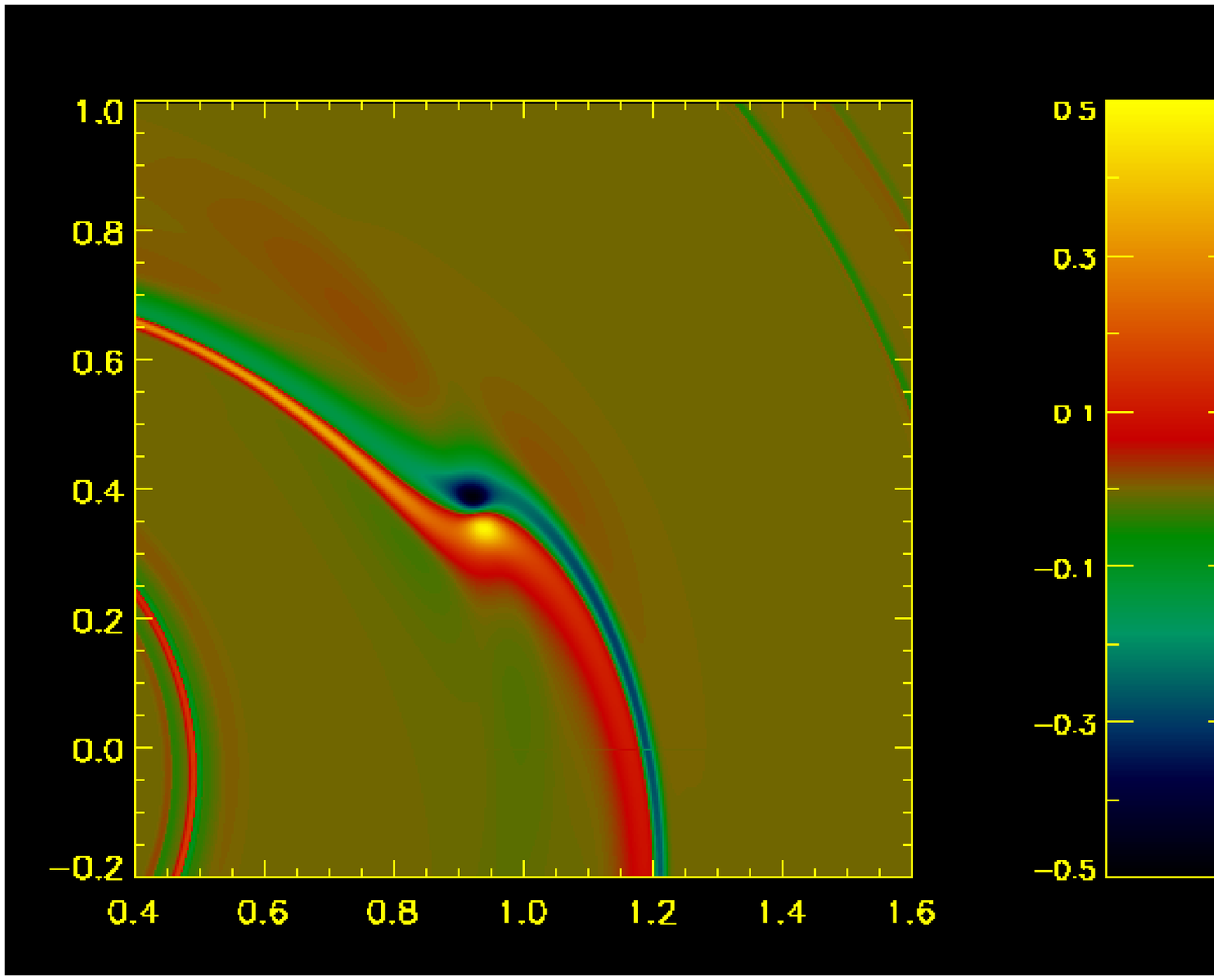}

  \caption{Simulations for the Neptune-mass planet in disk.
  \textit{Left panel}: (2a)Distribution of PV around the Neptune-mass
  planet at  $t = 70$ orbits for the run with resolution 800 x 3200.
  Shocks are well resolved. Vortensity is generated within the coorbital region,
  especially in the Roche lobe, where no shock exists. The azimuthal density gradient
  within the same region  acts as a source term of PV generation.
  \textit{Right panel}: (2b) Distribution of azimuthal density gradient
  at the same time.  Strong density gradient is observed within the Roche lobe and
  coorbital region.
} \label{fig2}
\end{figure}

\section{Summary and Discussion}
In this work, we carried out high resolution simulations on the
interaction between a protoplanetary disk and an embedded planet
with emphasis on the interplay between a disk and a planet under the
influence of baroclinic generation of PV. Our results are consistent
with classical analysis on the interaction between a protoplanetary
disk and an embedded planet through Lindblad torques. We confirmed
previous outcomes that non-axisymmetric RWI is likely to develop
under certain circumstances and have an important influence on the
migration of a planet inside an inviscid disk. We also found that
the generation of PV is more common and effective in disks with
non-barotropic EOS through the baroclinic instability, further
favoring the development of RWIs. As the asymmetry of the density
distribution induced by RWIs becomes prominent, the resulting
density blobs exert periodical and enhanced gravitational pull onto
the planet as they pass by the vicinity of the planet, which causes
the total torque received by the planet undergo large amplitude
oscillations.

Our analysis shows that strong vorticity has been generated around
the planet through baroclinic effect, this may help its core to
accrete materials faster in a way suggested in Klahr \& Bodenheimer
(2006) and shorten the time scale needed to form a Jupiter mass
planet. As a side effect of an inwardly migrating planet, RWIs
introduce non-axisymmetric density blobs along the way.  These
enhanced density blobs with strong vortices may help rapid formation
of new planet cores within them, especially in inner regions of
circumstellar disks where rapid precipitation and coagulation of
solid materials are likely to happen (Silverstone 2006). If these
new-born cores could survive during the migration of a giant planet
(Raymond 2006), they may produce Earth-like planets in the Habitable
Zones (Ji et al. 2007) or Hot Earths interior to a close-in giant
planet (Raymond 2006). One important issue is where this
non-barotropic mechanism is expected to occur in a protoplanetary
disk. Typically, at small radii ($r<10$ AU), the disk flow can not
dissipate its internal energy in a time less than the horseshoe
U-turn time, the locally isothermal approximation does not hold
there. Therefore, we expect this mechanism is likely to happen at
larger radii. On the other hand, the breakdown of the locally
isothermal approximation does not necessarily remove the temperature
variation. If temperature variation preserves at smaller radii, it
may still favor baroclinic effect. In realistic situations, this
could be a much more complicated process than what is described
here. Further investigations on these issues with full 3D
simulations and disk self-gravity (see also Baruteau \& Masset 2007;
Zhang et al. 2007 in this issue) are under way.

\begin{acknowledgments}
We thank the anonymous referee for useful comments and suggestions
that helped to improve the contents. J.H.J. acknowledges the
financial support by the National Natural Science Foundations of
China (Grants 10573040, 10673006, 10203005, 10233020) and the
Foundation of Minor Planets of Purple Mountain Observatory. S.O. was
partially supported by NSF grant AST-0407070.
\end{acknowledgments}

\end{document}